\documentclass[aps,prl, twocolumn, superscriptaddress,showpacs,lengthcheck]{revtex4-1}

\usepackage[T1]{fontenc}
\usepackage[latin1]{inputenc}
\usepackage[dvips]{graphicx,color}
\usepackage{rotating}
\usepackage{bm}        % \boldsymbol (bold math)
\usepackage{amssymb}   % \mathbb     (blackboard bold)
\usepackage{amsmath} 
\usepackage{amsfonts}
\def\jcp#1#2#3{J.~Chem.~Phys.~{\bf #1},\ #2\ (#3)}

\def\pra#1#2#3{Phys.~Rev.~A~{\bf #1},\ #2\ (#3)}
\def\prl#1#2#3{Phys.~Rev.~Lett.~{\bf #1},\ #2\ (#3)}

\def\k1{k_1}
\def\k2{k_2}
\def\q1{q_1}
\def\q2{q_2}

\def\({\left (}
\def\){\right )}
\def\[{\left [}
\def\]{\right ]}

\def\2b{\text{(2b)}}
\def\3b{\text{(3b)}}

\newcommand{\beq}{\begin{equation}}
\newcommand{\eeq}{\end{equation}}

\begin{document}
\date{\today}
\title{Three-body rf association of Efimov trimers}

\author{T. V. Tscherbul}
\affiliation{Harvard-MIT Center for Ultracold Atoms, Cambridge, Massachusetts 02138}
\affiliation{Institute for Theoretical Atomic, Molecular and Optical Physics,
Harvard-Smithsonian Center for Astrophysics, Cambridge, Massachusetts 02138}\email[]{tshcherb@cfa.harvard.edu}
\author{Seth T. Rittenhouse}
\affiliation{Institute for Theoretical Atomic, Molecular and Optical Physics, 
Harvard-Smithsonian Center for Astrophysics, Cambridge, Massachusetts 02138}

\begin{abstract}
We present a theoretical analysis of rf association of Efimov trimers in a 2-component Bose gas with short-range interactions. Using the adiabatic hyperspherical Green's function formalism to solve the quantum 3-body problem, we obtain universal expressions for 3-body rf association rates as a function of the $s$-wave scattering length $a$. We find that  the association rates scale as $a^{-2}$ in the limit of large $a$, and diverge as $a^3 a_{ad}^{3}$ whenever an Efimov state crosses the atom-dimer threshold (where $a_{ad}$ stands for the atom-dimer scattering length). Our calculations show that trimer formation rates as large as $\sim$10$^{-21}$ cm$^6$/s can be achieved with rf Rabi frequencies of order 1 MHz, suggesting that direct rf association is a powerful tool of making and probing few-body quantum states in ultracold atomic gases.
\end{abstract}


\maketitle

\clearpage
\newpage

The experimental study of universal phenomena in ultracold atomic gases \cite{BraatenHammer,Grimm06} has led to numerous recent advances in quantum few-body physics \cite{BraatenHammer}. In particular, weakly bound Efimov trimers formed by resonantly interacting bosons \cite{Grimm06} or fermions \cite{Lompe} provide a unified framework for understanding 3-body phenomena in atomic, nuclear, condensed-matter, and high-energy physics. The field of ultracold quantum gases provides an ideal playground for studying the Efimov physics because interactions of ultracold atoms can be tuned over many orders of magnitude by varying an external magnetic field near a Feshbach resonance \cite{Julienne}. Measurements of 3-body recombination rates as functions of the atom-atom scattering length $a$ have provided solid evidence for the existence and universal properties of the Efimov states in ultracold gases of  alkali-metal atoms \cite{Grimm06,Kh,Hulet,Zaccanti}, providing novel insights into the quantum dynamics of 3- and 4-body recombination \cite{Javier}.

While highly successful, experimental studies of Efimov trimers via 3-body loss measurements are limited to observations of zero-energy crossings of the trimers with the continuum \cite{Grimm06,Kh,Hulet}. Very recently, this limitation has been overcome by Lompe {\it et al.}  and later by Nakajima {\it et al.} \cite{Lompe}, who reported signatures of Efimov states in radiofrequency (rf) association spectra of an ultracold mixture of $^6$Li$_2$ dimers with $^6$Li atoms. These experiments have pioneered a radically new approach to few-body physics, in which Efimov states are ``actively'' created and probed via rf spectroscopy \cite{Multichann}, rather than ``passively'' observed via measurements of 3-body recombination rates. This approach is potentially very powerful, since rf spectroscopy is a well-established experimental tool for probing few- and many-body interactions in ultracold quantum gases \cite{Julienne,Ketterle}. Examples include measurements of binding  energies of Feshbach dimers using bound-bound and bound-free rf spectroscopy \cite{Julienne}, as well as many-body correlations and quantum phase transitions in ultracold quantum gases \cite{Ketterle}. Near-resonant rf fields can be used to create Feshbach molecules by rf association \cite{RF2body}, and manipulate magnetic Feshbach resonances in collisions of ultracold atoms \cite{Joerg,RFcontrol}.

In this Letter, we develop a quantum theory of 3-body rf association of Efimov trimers. By combining the adiabatic hyperspherical Green's function formalism \cite{Seth} with Fermi's Golden Rule, we derive the expressions for 3-body association rates as a function of the atom-atom scattering length $a$. 
Our results show that 3-body rf association rate scales as $a^{-2}$ in the limit of large $a$ and approaches infinity whenever an Efimov state crosses the atom-dimer threshold at positive $a$. We apply our theory to calculate the rate for the production of Efimov trimers by rf association in an ultracold gas of $^7$Li atoms. Our results show that large rf association rates can be achieved with Rabi frequencies available in the laboratory. Our theory can be generalized to describe rf-assisted 3-body phenomena in 3-component Fermi gases and atom-molecule mixtures \cite{Lompe, FeshbachChemistry}, providing a novel tool for studying rf association, spectroscopy, and chemical dynamics of universal quantum few-body states in ultracold atomic gases.

We begin by considering rf association of three free atoms $b$ + $b$ + $x$ leading to the formation of a weakly bound Efimov trimer $bbb$ where $b$ and $x$ refer to two different hyperfine states of a bosonic atom. The experimental prototype for such a system is a 2-component Bose gas of $^7$Li atoms in their lowest hyperfine states \cite{Kh,Hulet}. The system of three particles is characterized by two $s$-wave scattering lengths $a_{bb}$ and $a_{bx}$. Throughout this work, we assume that $a_{bx}=0$, and examine the dependence of the scattering observables on a single parameter -- the scattering length $a=a_{bb}$. 
The thermal rate constant for rf association of three distinguishable atoms into a weakly bound trimer can be written in the form
\begin{equation}\label{K3}
K_{3} = \frac{\hbar k}{\mu}\sigma_{if} = \frac{\hbar k}{\mu} \frac{32\pi^2}{k^5}|S_{if}(\epsilon,\Omega,\Delta)|^2
\end{equation}
where $\sigma_{if}$ is the cross section for stimulated  radiative association, $\mu$ is the 3-body reduced mass, $k^2=2\mu \epsilon/\hbar^2$, $\epsilon$ is the collision energy, $\Omega$ the Rabi frequency of the rf field, $\Delta = \hbar\omega - \epsilon_f$ is the detuning from resonance, and $\epsilon_f$ is the binding energy of the Efimov trimer formed by rf association. While (\ref{K3}) is similar in form to the expressions encountered in the theory of photoassociation lineshapes  \cite{Napolitano,Cote}, the important prefactor $32\pi^2/k^5$ have been introduced to properly account for the 3-body phase space density \cite{Esry}. In deriving Eq. (\ref{K3}), we assumed that the temperature of the ultracold gas is sufficiently low so that only $s$-wave collisions contribute to the cross section. It is convenient to parametrize the $S$-matrix element by a Lorentzian \cite{Napolitano,Cote}
\begin{equation}\label{Smatrix}
|S_{if}(\epsilon,\Delta,\Omega)|^2 = \frac{\gamma_d\gamma_s(\epsilon,\Omega)}{(\epsilon-\Delta)^2 + (\gamma/2)^2},
\end{equation}
where $\gamma_s(\epsilon,\Omega)$ is the stimulated width,  $\gamma_d$ is the decay width introduced to account for non-radiative decay of Efimov trimers via 3-body recombination \cite{Note}, and $\gamma = \gamma_s+\gamma_d$ is the total width. The stimulated width is given by the Fermi's Golden Rule \cite{Napolitano,Cote} 
\begin{equation}\label{gamma_s}
\gamma_s(\epsilon,\Omega) = 2\pi \frac{\hbar^2 \Omega^2}{4} |\langle \Psi_i | \Psi_f\rangle |^2 
\end{equation}
where $\Omega = 2\langle x | \hat{\mu} | b \rangle /\hbar$ is the atomic Rabi frequency, $\hat{\mu}$ is the transition dipole moment operator, $\Psi_i$ is the wavefunction for the initial 3-body continuum state, and $\Psi_f$ is the wavefunction of the Efimov trimer. In the limit of zero collision energy, $|S_{if}|^2 \sim \epsilon^2$ \cite{Macek}, and we can define an energy-independent normalized Franck-Condon factor (FCF)
\begin{equation}\label{FCF0}
\mathcal{F}_{if} =  |\langle \Psi_i | \Psi_f\rangle |^2 / \epsilon^{2},
\end{equation}
Substituting this expression in Eq. (\ref{K3}), we obtain for the 3-body association rate on resonance ($\epsilon=\Delta$)
\begin{equation}\label{K3_FGR}
K_{3} =  \frac{16\pi^3 \hbar^5}{\mu^3} (\hbar \Omega)^2 \frac{\mathcal{F}_{if}}{\gamma_d} 
\end{equation}
In deriving this expression, we made the assumption $\gamma_s\ll \gamma_d$, which is well justified for short-lived Efimov states with observed decay widths on the order of several MHz \cite{Kh,Lompe}. In order to evaluate the FCF (\ref{FCF0}), we use the adiabatic hyperspherical Green's function formalism \cite{Seth}. We assume that the initial and final 3-body wavefunctions take the form $\Psi_\beta = R^{-5/2}F_\beta (R) \Phi_\beta(R;\omega)$, where $R$ is the hyperradius, $\omega$ are the hyperangles, and $\beta=i,f$. Here $\Phi_i$ is assumed to the hyperangular channel function which corresponds to three free particles in the $R \gg a $ limit, and $\Phi_f$ is taken to be the channel function which supports an Efimov trimer state. Substituting these wavefunctions to Eq. (\ref{FCF0}), we find
\begin{equation}\label{FCF}
\mathcal{F}_{if} = \epsilon^{-2} \left[\int F_i(R) W_{if}(R)F_f(R) dR\right]^2,
\end{equation}
where $W_{if}(R)$ is the hyperangular channel function overlap at a fixed hyperradius and the radial functions $F_\beta(R)$ are solutions to the adiabatic Schr{\"o}dinger equations \cite{BraatenHammer,Seth}
\begin{equation}\label{radial_eqs}
-\frac{1}{2\mu} \left( \frac{\partial^2}{\partial R^2} - U_\beta(R)\right) F_{\beta} (R) = \epsilon_\beta F_\beta(R),
\end{equation}
subject to the boundary condition $F_\beta (R_0) = 0$, where $R_0$ is the radial cutoff parameter related to the experimentally relevant 3-body parameter \cite{BraatenHammer} by $R_0 \approx 0.22 a_+$ where $a_+$ is the position of the first universal minimum in the 3-body recombination rate of three identical bosons. At $R\to\infty$, the energy-normalized scattering wavefunction takes the form $F_i(R) \simeq (2\mu/ \pi \hbar^2 k)^{1/2} \sin (kR+\delta)$ \cite{Julienne}, where $\delta$ is the scattering phase shift. For the bound-state wavefunction, we impose the standard boundary condition $F_f(\infty) = 0$. In Eq. (\ref{radial_eqs}), $U_\beta(R) = [(\nu_\beta(R) +2)^2-1/4]/R^2+Q(R)$ are the adiabatic hyperspherical potentials, $\nu_\beta(R)$ are the adiabatic eigenvalues and $Q(R)$ are the diagonal non-adiabatic corrections \cite{Seth}. The angular integral $W_{if}$ in Eq. (\ref{FCF}) can be evaluated by expanding the functions $\Phi_\beta(R;\omega)$ in hyperspherical harmonics \cite{BraatenHammer,Seth}.

\begin{figure}[t]
	\centering
	\includegraphics[width=0.47\textwidth, trim = 0 0 0 0]{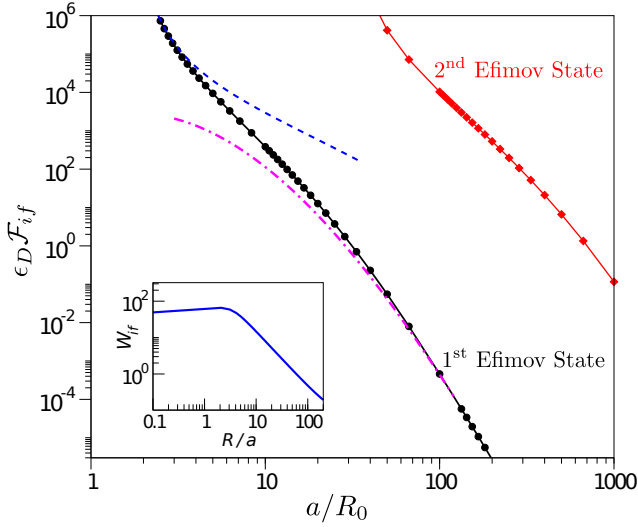}
	\renewcommand{\figurename}{Fig.}
	\caption{The FCF (\ref{FCF}) in units of $\epsilon_D^{-3}$ as a function of the reduced scattering length $a/R_0$. Symbols -- numerical calculations using Eq. (\ref{FCF}); dashed lines -- analytic results obtained for  $a\to 0$ (blue dashed line) and $a\to \infty$ (magenta dot-dashed line). The inset shows the radial dependence of the angular integral $W_{if}(R)$ in reduced units.}\label{fig:FCF}
\end{figure}

Figure \ref{fig:FCF} shows the calculated FCFs for the first two Efimov states in unites of the dimer binding energy $\epsilon_D^{-3}$ ($\epsilon_D=\hbar^2/ma^2$) as a function of the atom-atom scattering length in units of the 3-body cutoff $R_0$.  As $a/R_0$ increases, the Efimov states become more strongly bound, and the bound-state wavefunction $\Psi_f(R)$ in Eq. (\ref{FCF}) becomes more localized in the small $R$ region, where the amplitude of the scattering wavefunction $\Psi_i(R)$  is suppressed due to a steep increase of the adiabatic potential $U_i(R)$ \cite{Seth}. The angular overlap integral $W_{if}(R)$ becomes constant in the limit $a\to 
\infty$ (see the inset in Fig. \ref{fig:FCF}), so the large-$a$ behavior of the FC overlap is determined solely by the wavefunctions in Eq. (\ref{FCF}). The decrease of the FCF with $a$ shown in Fig. \ref{fig:FCF} is therefore a consequence of diminishing overlap between the initial and final wavefunctions.

The FCFs for different Efimov states have similar shapes, decreasing monotonously with increasing $a/R_0$. The FCF for the first excited Efimov state is shifted toward higher $a$ by the universal factor of 22.7 \cite{BraatenHammer}.  Since the FCFs scale as $\epsilon_D^{-3}$ we can expect that the FCFs for different Efimov states will be related by a factor of $22.7^6$.  However, in general the decay width of consecutive Efimov states will decrease by the universal scale factor of $22.7^2$ meaning that the zero temperature 3-body association rate constant in Eq. (\ref{K3_FGR}) will scale as $22.7^4$ between neighboring Efimov states.

 To make these arguments more quantitative, let us first consider the limit of small $a$, where the trimer state approaches the atom-dimer threshold and the radial wavefunction is well approximated by 
\begin{equation}
F_{f}\left(  R\right)  \approx\sqrt[4]{4\dfrac{2\mu}{\hbar^{2}}\left(
\epsilon_f-\epsilon_D\right)  } \exp\left(  \sqrt{2\mu\left(  \epsilon_f-\epsilon_D\right)
}R\right)
\end{equation}
In this limit, most of the overlap occurs when $R\gg a$ and thus we approximate the initial energy-normalized wavefunction as $F_{i}\left(  R\right)  =\sqrt{\dfrac{\mu}{\hbar^{2}}}\sqrt{R}J_{2}\left(kR\right)$, where $J_2$ is a Bessel function. Examining the numerical results shown in the inset of Fig. \ref{fig:FCF} yields $W\left(  R\right) \rightarrow {C}a^{3/2}/R^{3/2}$ with $C =452.741$. Using this result in Eq.~(\ref{FCF}), we find
\begin{equation}\label{a_large}
\mathcal{F}_{if} =\sqrt{2}
C^2\left(  \dfrac{a}{\hbar}\right)^{3}
\sqrt{\mu^3 \varepsilon}
\dfrac{\left(2\varepsilon+\epsilon-2\sqrt{\varepsilon\left(\varepsilon+\epsilon\right)  }\right)^2}{\epsilon^4}
\end{equation}
where $\varepsilon=\epsilon_f-\epsilon_D$ is the binding energy of the Efimov state relative to the atom-dimer threshold. 
 The analytical result shown in Fig. \ref{fig:FCF} is in excellent agreement with the full numerical calculation; the difference does not exceed 10\% for $a/R_0<2.5$.  In the limit where $ka \ll 1$ and $\epsilon \ll \varepsilon$ Eq. (\ref{a_large}) takes on the simple universal form $\mathcal{F}_{if} \rightarrow \mu^3C^2a^3a_{ad}^3/\left(\sqrt{2}3^{3/4} \hbar^6\right)$ where $a_{ad}=\sqrt{3\hbar^2/4m\varepsilon}$ is the scattering length of a bosonic atom scattering of a boson-boson Feshbach molecule. As the first Efimov state crosses the atom-dimer threshold and becomes unbound at $a/R_0 \sim 2.5$, $a_{ad}$ tends to infinity and $\mathcal{F}_{if}$ diverges as shown in Fig. \ref{fig:FCF}.

We now consider the opposite limit $a\to \infty$, in which the Efimov state becomes strongly bound relative to the atom-dimer threshold and the final wavefunction is only non-zero in the region $R<a$. Further, the adiabatic potentials in Eq. (\ref{radial_eqs}) become universal when $R \ll a$, giving the initial and final wavefunctions
\begin{align}\notag
F_{i}\left(  R\right)   &\sim\sqrt{\dfrac{\mu}{\hbar^2}} \sqrt{R}J_{\nu
+2}\left(  kR\right), \\
F_{f}  &=\sqrt{\dfrac{4E\mu\sinh\left(  \pi s_{0}\right)  }{\pi\hbar
^{2}s_{0}}}\sqrt{R}K_{is_{0}}\left(  \sqrt{2\mu \epsilon_f}R\right),
\end{align}
where $K_{is_0}$ is a modified Bessel function of the second kind of imaginary order and $s_0^2 = 1.01251$ \cite{BraatenHammer,Seth}.
As shown in the inset of Fig. \ref{fig:FCF},  $W\left(  R\right)  \rightarrow B_{\nu}$ with  $B_1=47.898$ in the limit $a\rightarrow\infty$. Substitution of this result in Eq. (\ref{FCF}) gives
\begin{multline}\label{a_small2}
\mathcal{F}_{if} = \dfrac{B_{\nu}^2}{2}\left(  \dfrac{\epsilon}{\epsilon_f}\right)  ^{\nu}\dfrac
{\sinh\left(  \pi s_{0}\right)  }{\epsilon_f^3  s_{0}}\dfrac{\Gamma^2\left( \nu_- \right)  \Gamma^2\left( \nu_+ \right)  }
{\pi^2\Gamma^2\left(  \nu+3\right)  } \\
  \times \left[{}_2F_{1}\left(  \nu_-, \nu_+ ;\nu+3;-\dfrac{\epsilon}{\epsilon_f}\right)\right]^2,
\end{multline}
where $\nu_\pm =  (\nu+4\pm is_{0})/2$ and ${}_2F_{1}$ is a hypergeometric function.
This expression for the FC overlap is correct in the case where $|ka| \gg 1$ not only for the $\nu = 1$ initial channel of interest in this work, but also for all incident $s$-wave channels with $\nu = -1,1,3,5,...$. 
By replacing $\epsilon_f$ with $\epsilon_{fn}=\epsilon_{f}\left(22.7\right)  ^{2n}$ we can get the overlap of the $n$th Efimov state in terms of the first with binding $\epsilon_{f}$. In the large $a$ limit, but where $ka\ll1$, there is extra suppression of the wavefunction in the $\nu = 1$ incident channel due to the additional tunneling in the $a<R<1/k$ regime producing and extra factor of $\left( 6/ ka\right) ^{2}$ in the FCF. When $a \rightarrow \infty$ and in the $ka \ll 1$ regime the FC overlap takes on the simple universal form of $\mathcal{F}_{if}\rightarrow 1661.12 \hbar^2/\mu a^2 \epsilon_f^4$.  It should be noted that  this large $a$ behavior was found under the assumption that the initial 3-body state is described by a single hyperangular channel function. In truth, in the region where $R \ll a$ the initial state of the system is multi-channel in nature, and a more complete description will be required. Fortunately the FCFs found in Eq. (\ref{a_small2}) are fully generalized for overlaps in this regime.  Incorporating the full multi-channel nature of the system is the subject of ongoing study.

\begin{figure}[t]
	\centering
	\includegraphics[width=0.4\textwidth, trim = 0 0 0 0]{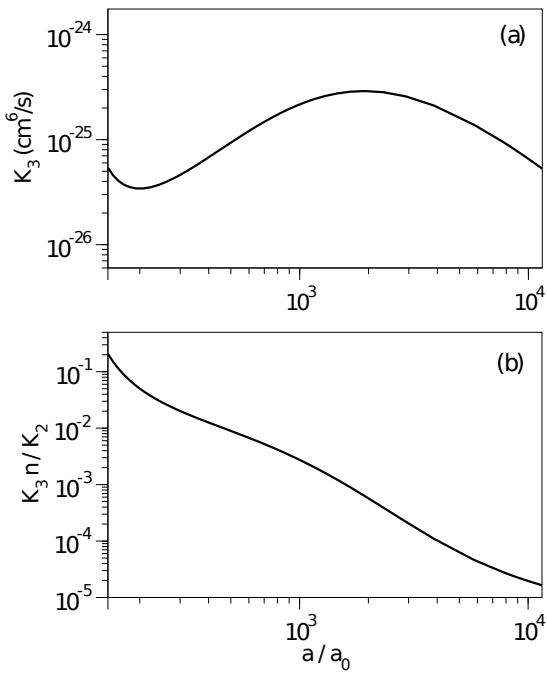}
	\renewcommand{\figurename}{Fig.}
	\caption{(a) The 3-body rf association rate for $^7$Li as a function of $a$ calculated for $\Omega/2\pi = 10$ kHz and $R_0=58a_0$ \cite{Kh}. (b) The relative efficiency of 3-body vs 2-body rf association ($K_3n/K_2$) plotted as a function of $a$ for $n = 10^{12}$ cm$^{-3}$ and $T=1\mu K$. }\label{fig:K3}
\end{figure}

As an application of the universal theory developed above, we evaluate the rate for the production of Efimov trimers by rf association in an ultracold gas of $^7$Li atoms. The Efimov  physics in this system has been studied in detail experimentally \cite{Kh,Hulet}. We assume that the $^7$Li atoms are prepared in two lowest hyperfine states $|b\rangle = |11\rangle$ and $|x\rangle = |10\rangle$ and the dc magnetic field is tuned such that $a_{bx}=0$. Figure \ref{fig:K3}(a) shows the 3-body rf association rate as a function of $a$ calculated using typical experimental parameters $\Omega/2\pi = 10$ kHz \cite{Lompe}, $R_0 = 58a_0$, and $\gamma_d=4.5$ MHz \cite{Kh,TBP}. The association rate takes a minimum value of $3.3\times 10^{-26}$ cm$^3$/s at $a= 200 a_0$ and a maximum value of $2.9\times 10^{-25}$ cm$^3$/s at $a= 1940 a_0$. These rate constants are large enough as to be easily observable experimentally via monitoring atom loss from an optical dipole trap \cite{Kh}.  The oscillating behavior of $K_3$ observed at finite $a$ is in marked contrast with the familiar $a^4$ scaling of the 3-body recombination rate \cite{BraatenHammer}.

As the trimer binding energy approaches the atom-dimer threshold, it is natural to expect that rf association will start producing not only Efimov trimers, but also weakly bound dimers. In order to examine the interplay between these processes, we evaluated the 2-body rf association rate $K_2$ using the Fermi's Golden Rule \cite{Julienne,Cote} in the limit $ka\ll 1$ \cite{TBP} with the result
\begin{equation}\label{K2}
K_2 = \frac{4\sqrt{2}\pi^{3/2}e^{-1/2}}{\hbar} \frac{(\hbar\Omega)^2}{k_BT} a^3
\end{equation}
In Fig. \ref{fig:K3}(b), we plot the ratio $g = K_3n/K_2$, which quantifies the relative efficiency of 2-body versus 3-body rf association, for typical experimental conditions $n=10^{12}$ cm$^{-3}$ and $T = 1\,\mu$K \cite{Kh} (note that $g$ is independent of $\Omega)$. The ratio declines monotonously with $a$ due to the $a^3$ scaling behavior of the 2-body association rate (\ref{K2}). The results shown in Fig. \ref{fig:K3} thus suggest that 3-body rf association experiments should be performed at small-to-moderate values of $a$ to suppress 2-body rf association.

In summary, we have developed a universal theory of rf association of Efimov trimers in a 2-component Bose gas with short-range interactions. We have applied the theory to evaluate the rates for 3-body rf association in an ultracold gas of $^7$Li atoms and derived analytical expressions for the association rates, which are valid in the limits of large $a$  ($K\sim a^{-2}$) or infinitely weakly bound Efimov states ($K\sim a^3a_{ad}^3$).  We find that rf association of weakly bound (atom-dimer-like) Efimov trimers is more efficient than that of strongly bound Efimov trimers (Fig.~\ref{fig:K3}), and rationalize this result based on simple overlap arguments. We also examine the competing process of 2-body rf association, and find that  the relative efficiency of 3-body versus 2-body rf association decreases with increasing $a$ (Fig.~\ref{fig:K3}b).
The methods developed in this work can easily be generalized to describe rf-induced few-body phenomena in 3-component Fermi gases \cite{Lompe}, atom-dimer mixtures \cite{FeshbachChemistry} and 4-body systems \cite{Javier} or in 3-body systems with multi-channel 2-body interactions \cite{Multichann, Macekmc}. These methods might also be applied to isotropically trapped systems where a transition matrix element to an excited trap state can be calculated in stead of a rate constant.

Our findings provide quantitative insight into the mechanisms of rf association of universal trimer states in ultracold quantum gases, and may thus have significant implications for research in this rapidly expanding area of physics \cite{BraatenHammer,Grimm06,Kh,Hulet,Lompe}. We have shown that rf association of ultracold Li atoms into Efimov trimers occurs at a substantial rate of $\sim$10$^{-25}$ cm$^6$/s (Fig. \ref{fig:K3}) for Rabi frequencies of order 10 kHz, which can be readily achieved in the laboratory \cite{Lompe,RFcontrol}. The 3-body association rate may be further increased by raising the Rabi frequency, since $K_3\sim \Omega^2$ (\ref{K3_FGR}), so values of $K_3$ on the order $\sim10^{-21}$ cm$^{6}$/s should be readily achievable with $\Omega/2\pi \sim 1$ MHz. Such Rabi frequencies can be generated in atom chip experiments \cite{Joerg}, which would provide an ideal setting for creating large quantities of Efimov trimers, heretofore an unsolved problem due to their large decay rates \cite{Lompe}. The results shown in Fig. \ref{fig:K3} suggest that rf association experiments in Li should be performed at small-to-moderate values of $a$ (between 200 and 2000$a_0$) to avoid the competing processes of 2-body rf association and 3-body recombination.

This work was supported by NSF grants to the Harvard-MIT CUA and ITAMP at Harvard University and the Smithsonian Astrophysical Observatory.


\begin{thebibliography}{99}
\bibitem{Note}
We assume that $\gamma_d$ for a given Efimov trimer is independent of $\epsilon_f$.
\end{thebibliography}
\end{document}